\documentclass[floatfix,showpacs]{revtex4}
\usepackage{epsfig}
\usepackage{graphicx}
\usepackage[german, english]{babel}
\usepackage{amsmath}
\usepackage{amssymb}
\usepackage{ifthen}
\usepackage{epsfig}
\newcounter{fig}   \newcommand{\lbfig}[1]{\refstepcounter{fig}
\label{#1} } 
\newcommand{\vphi}{\varphi}
\newcommand{\Tr}{{\rm Tr}}

\begin{document}

\title{ Electromagnetic Interaction in the System of Multimonopoles\\ 
and Vortex Rings}

\author{Yasha Shnir} 

\affiliation{Institut f\"ur Physik, Universit\"at Oldenburg,
D-26111, Oldenburg, Germany}


\begin{abstract}
Behavior of static axially symmetric 
monopole-antimonopole and vortex ring solutions of the 
$SU(2)$ Yang-Mills-Higgs theory in an external uniform magnetic field 
is considered. It is argued that the axially symmetric monopole-antimonopole 
chains and vortex rings can be treated as a bounded 
electromagnetic system of the magnetic charges and the   
electric current rings. 
The magnitude of the external field is a parameter which may be used 
to test the structure of the static  
potential of the effective 
electromagnetic interaction between the monopoles with 
opposite orientation in the group space.  
It is shown that for a non-BPS solutions 
there is a local minimum of this potential. 
\end{abstract}


\maketitle


\section{Introduction}
The structure of the vacuum of the $SU(2)$ Yang-Mills-Higgs (YMH) theory is 
rather nontrivial (see. e.g., \cite{Manton-book,S-book}), 
there are  spherically symmetric monopoles with unit 
topological charge \cite{mono}, 
axially symmetrical multimonopoles of higher topological charge 
\cite{WeinbergGuth,RebbiRossi,mmono}, and solutions with platonic symmetry  
\cite{Manton-book,monoDS}. There is also a monopole-antimonopole (M-A) pair  
static solution, 
which is a deformation of the topologically trivial sector  \cite{Rueber,mapKK}. 
Recently, another deformations of this sector, which
represented monopole-antimonopole 
chains and vortex rings, were discussed \cite{KKS}. 
There are also electrically charged generalizations of these solutions
\cite{jul,wein,dyonhkk}, which appear due to excitation of the gauge zero modes  
of the underlying electrically neutral solutions. 

These solutions are characterized by 
two integers, the winding number $m$ in polar angle $\theta$ 
and the winding number $n$ in azimuthal angle $\varphi$. 
The structure of the nodes of the Higgs field depends on the values of these integers, 
there are both chains of zeros and rings. However, only the winding number 
$n$ has a meaning of the topological charge of the configuration. 

In the Bogomol'nyi-Prasad-Sommerfield (BPS) limit of vanishing Higgs potential
the spherically symmetric monopole and axially symmetric multimonopole solutions, 
which satisfy the first order Bogomol'nyi equations \cite{BPS}
as well as the second order field equations,
are known analytically \cite{mmono}. 

Taubes proved that in the $SU(2)$ YMH theory
a smooth, finite energy magnetic dipole solution of the second order
field equations, which do not satisfy the Bogomol'nyi equations,
could exist \cite{Taubes}. In his consideration the space of the field 
configurations and the energy functional are considered as the manifold and the function, 
respectively. For a monopole-antimonopole pair (M-A) the map $S^2 \to S^2$ 
has a degree zero, thus it is a deformation of the topologically trivial sector. A generator 
for the corresponding homotopy group is a non-contractible loop which describes creation of a 
monopole-antimonopole pair with relative orientation in the isospace $\delta = -\pi$ 
from the vacuum, 
separation of the pair, rotation of the monopole by 
$2\pi$ and annihilation of the pair back into vacuum.
Minimization of the energy functional 
along such a loop yields an equilibrium state in the middle of the loop where the monopole
is rotated by $\pi$ 
and $\delta = 0$. 
Such an axially symmetric configuration, with 
two zeros of the  Higgs field located symmetrically on the positive and negative 
$z$-axis, corresponds to a saddlepoint 
of the energy functional, a monopole and antimonopole in static equilibrium, 
a magnetic dipole \cite{Rueber,mapKK}. 

It is interesting to compare the situation with the case of two monopoles. 
It was shown long ago \cite{Manton77}, that 
there is a balance of two 
long-range interactions between the BPS monopoles, which are 
mediated by the massless photon and the massless scalar 
particle, respectively. The balance of interactions, which yields the magnetic dipole solution,
is a bit more subtle. 
Indeed, a monopole and an antimonopole can only be 
in static equilibrium, if they are close enough to experience a
repulsive force \cite{mapKK,KKS} which may balance the electromagnetic and scalar attractions.  
In this case we have a complicated pattern 
of short-range interactions and the structure of the nodes of the Higgs fields 
strongly depends on the scalar coupling \cite{KKS}. All these Yukawa interactions result in an 
effective potential of interaction whose local minimum corresponds, for example, 
to the static dipole solution.   

Evidently, an external electromagnetic field may be used as a ``dipstick'' to test 
the structure of the net potential of the 
interaction between the poles. In this note I argue that one can make use of  
an effective electromagnetic interaction in such solutions, 
which encompasses the complicated picture 
of the short-range Yukawa forces above. Further, to support this interpretation, I 
study the interaction 
of different static axially symmetric deformations of the topologically trivial sector, both monopoles 
and vortex rings, 
with an external uniform magnetic field.

In section II we present the action, the axially
symmetric Ansatz and the boundary conditions.
In section III we discuss 
the effect of coupling of the system with an external homogeneous magnetic field and consider  
the behavior of the forced monopole-antimonopole pairs and vortex rings.
We present our conclusions in section IV.

\section{\bf Yang-Mills-Higgs solutions: Monopole-antimonopole chains and vortex rings}

The Lagrangian of the Yang-Mills-Higgs model 
is given by 
\begin{equation}
-L_0 = \int\left\{ \frac{1}{2} \Tr\left( F_{\mu\nu} F^{\mu\nu}\right)
                +\frac{1}{4} \Tr\left( D_\mu \Phi D^\mu \Phi \right)
		+\frac{\lambda}{8} \Tr\left[ \left(\Phi^2 - \eta^2\right)^2 \right] 
	\right\} d^3 r \ ,
\label{lag}
\end{equation}
with 
$su(2)$ gauge potential $A_\mu = A_\mu^a \tau^a/2$,
field strength tensor
$
F_{\mu\nu} = \partial_\mu A_\nu - \partial_\nu A_\mu + i e [A_\mu, A_\nu]$, 
and covariant derivative of the Higgs field $\Phi = \phi^a \tau^a$ in the adjoint representation
$
D_\mu \Phi = \partial_\mu \Phi +i e [A_\mu, \Phi]$. 
Here $e$ denotes the gauge coupling constant, $\eta$ the vacuum expectation value of the
Higgs field and $\lambda$ the strength of the Higgs selfcoupling. 

The static regular solutions of the corresponding field equations were constructed numerically 
by employing of the axially symmetric Ansatz \cite{KKS}
for the gauge and the Higgs fields 
\begin{eqnarray}
A_\mu dx^\mu
& = &
\left( \frac{K_1}{r} dr + (1-K_2)d\theta\right)\frac{\tau_\vphi^{(n)}}{2e}
-n \sin\theta \left( K_3\frac{\tau_r^{(n,m)}}{2e}
                     +(1-K_4)\frac{\tau_\theta^{(n,m)}}{2e}\right) d\vphi
\label{ansatzA}\nonumber \\
\Phi
& = &
\Phi_1\tau_r^{(n,m)}+ \Phi_2\tau_\theta^{(n,m)} \  .
\label{ansatzPhi}
\end{eqnarray}
The Ansatz is written in the basis of $su(2)$ matrices $\tau_r^{(n,m)},\tau_\theta^{(n,m)} $ and 
$\tau_\vphi^{(n)}$ which are defined as the dot product of the Cartesian vector of Pauli 
matrices $\vec \tau $ and the spacial unit vectors   
\begin{eqnarray}
{\hat e}_r^{(n,m)} & = & \left(
\sin(m\theta) \cos(n\vphi), \sin(m\theta)\sin(n\vphi), \cos(m\theta)
\right)\ , \nonumber \\
{\hat e}_\theta^{(n,m)} & = & \left(
\cos(m\theta) \cos(n\vphi), \cos(m\theta)\sin(n\vphi), -\sin(m\theta)
\right)\ , \nonumber \\
{\hat e}_\vphi^{(n)} & = & \left( -\sin(n\vphi), \cos(n\vphi), 0 \right)\ ,
\label{unit_e}
\end{eqnarray}
respectively. The gauge field functions $K_i$, $i=1,\dots,4$ and two Higgs field functions 
$\Phi_1, \Phi_2$ depend on the coordinates $r$ and $\theta$. The axial symmetry of the configuration 
allows also to consider the system in cylindrical coordinates with the planar radius 
$\rho = r\sin \theta$.

The Ansatz (\ref{ansatzA}) is axially symmetric 
in a sense that a spacial rotation around the $z$-axis 
can be compensated by Abelian gauge transformation
$U = \exp \{i\omega(r,\theta) \tau_\vphi^{(n)}/2\}$ which leaves the Ansatz form-invariant. However the 
gauge potential and the scalar field transform as 
$$
A_\mu^\prime = U A_\mu U^\dagger +\frac{i}{e}(\partial_\mu U)U^\dagger, \qquad 
\Phi^\prime = U\Phi U^\dagger 
$$
respectively. Then the  structure functions of the Ansatz transforming as 
\cite{mapKK}
{\begin{small}
\begin{equation} \label{gauge-rot}
\begin{split}
K_1 &\to K_1 - r \partial_r \omega\, ;\quad 
K_2 \to K_2 + \partial_\theta \omega\, ;\\
\left(K_3 + \frac{\cos (m\theta)}{\sin\theta}\right) &\to 
\left(K_3 + \frac{\cos (m\theta)}{\sin\theta}\right)\cos \omega + 
\left(1-K_4- \frac{\sin (m\theta)}{\sin\theta}\right)\sin\omega\, ;\\
\left(1-K_4- \frac{\sin (m\theta)}{\sin\theta}\right) &\to 
-\left(K_3 + \frac{\cos (m\theta)}{\sin\theta}\right)\sin \omega + 
\left(1-K_4- \frac{\sin (m\theta)}{\sin\theta}\right)\cos\omega\, ;\\
\Phi_1 &\to \Phi_1 \cos \omega + \Phi_2 \sin \omega \, ; \quad 
\Phi_2 \to -\Phi_1 \sin \omega + \Phi_2 \cos \omega \, . 
\end{split}
\end{equation}
\end{small}
}
~
To obtain a regular solution we make use of the $U(1)$ gauge symmetry 
to fix the gauge \cite{KKT}. 
We impose the condition 
$$
G_f = \frac{1}{r^2}\left(r\partial_r K_1 - \partial_\theta K_2\right) = 0 \, .
$$

The regular solutions with finite energy density and correct asymptotic 
behavior are constructed numerically by imposing the 
boundary conditions. The regularity of the energy density functional  
at the origin requires 
$$
K_1(0,\theta)=0\ , \ \ \ \ K_2(0,\theta)= 1 \ , \ \ \ \
K_3(0,\theta)=0 \ , \ \ \ \ K_4(0,\theta)=1 \ , \ \ \ \
$$
$$
\sin(k\theta) \Phi_1(0,\theta) + \cos(k\theta) \Phi_2(0,\theta) = 0 \ ,
$$
$$
\left.\partial_r\left[\cos(k\theta) \Phi_1(r,\theta)
              - \sin(k\theta) \Phi_2(r,\theta)\right] \right|_{r=0} = 0 
$$
that is $\Phi_\rho(0,\theta) =0$ and $\partial_r \Phi_z(0,\theta) =0$.

Since here we are discussing only 
the topologically trivial sector of the model, the related configurations 
at infinity required to tend to a pure gauge 
\begin{equation} 
\label{rand}
\Phi \ \longrightarrow U \tau_z U^\dagger \   , \ \ \
A_\mu \ \longrightarrow  \ i \partial_\mu U U^\dagger \ ,
\end{equation}
where  $U = \exp\{-i k \theta\tau_\vphi^{(n)}\}$ and $m=2k$.  Therefore 
in terms of the functions $K_1 - K_4$, $\Phi_1$, $\Phi_2$ these boundary
conditions read \footnote{In earliest study of the axially symmetric configurations, we worked in the
gauge where the scalar field components on the spacial asymptotic do not depend 
on the polar angle \cite{mapKK,KKS}.}
\begin{equation}
K_1 \longrightarrow 0 \ , \quad
K_2 \longrightarrow 1 - 2k \ , \quad
K_3 \longrightarrow 0 \ , \quad 
K_4 \longrightarrow 1- \frac{2\sin(k\theta)}{\sin\theta} \ ,
\label{K4infty}
\end{equation}
\begin{equation}        \label{Phiinfty}
\Phi_1\longrightarrow  \cos(k\theta) \ , \ \ \ \ \Phi_2 \longrightarrow \sin(k\theta) \ .
\end{equation}
Regularity on the $z$-axis, finally, requires
$$
K_1 = K_3 = \Phi_2 =0 \ , \ \ \  \
\partial_\theta K_2 = \partial_\theta K_4 = \partial_\theta \Phi_1 =0 \ ,
$$
for $\theta = 0$ and $\theta = \pi$.

Thus, the deformations of the topologically 
trivial sector are classified according to the values of the winding numbers 
$k$ and $n$. The branch 
of solutions with $k=1$ corresponds to the monopole-antimonopole pair M-A $(n=1)$ \cite{mapKK}, 
the charge-2 
monopole-antimonopole pair $(n=2)$ \cite{KKS,PT} and a single vortex ring $(n \ge 3)$. Another branch 
with $k=2$ corresponds to the monopole-antimonopole chain M-A-M-A 
with 4 nodes of the Higgs field on the $z$-axis 
 $(n=1)$, chain of 4 double nodes  $(n=2)$ and the system of two vortex rings  $(n\ge 3)$  \cite{mapKK}.
The energy of these configurations increases with $n$.

The axially symmetric solutions under consideration are characterised by a non-vanishing magnetic 
dipole moment \cite{mapKK,KKS}. 
It can be read off from the asymptotic form of the gauge field at infinity  
\begin{equation}
A_\mu dx^\mu = {\mu}\frac{\sin^2\theta}{2r}\tau_z d\vphi \ , 
\label{magmom}
\end{equation}
where $\mu$ is a (dimensionless) magnetic dipole moment.

Note that  
the pattern of interaction between the monopoles is very different from a naive
picture of electromagnetic interaction of point-like charges \cite{Manton77,Manton-book,S-book}.
Indeed, there is such an 
attractive force between well separated monopole and antimonopole,  
in the singular gauge, it is mediated by the $A^3$ component of the vector field.  
However, this field is massless only outside of the monopole core. On the other 
hand, it is known that the BPS monopoles do not interact at any separation which 
allows us to make use of the powerful moduli space approach 
(see, e.g., \cite{Manton-book}). The reason 
is that the repulsive vector interaction between the poles is always balanced by 
the scalar interaction. 

However, the axially symmetric configurations which we are discussing, 
are not solutions of the 
first order Bogomol'nyi equation, even in the limit
of vanishing scalar coupling where scalar interaction also becomes long-range.  
They are deformation of the topologically trivial sector and, on the 
scale of characteristic size of the axially 
symmetric solutions,   
both the scalar particle and the $A_\mu^3$ vector boson
remain massive. Furthermore,
the vector bosons $A_\mu^\pm$ also mediate the short-range Yukawa interactions 
between the monopoles and we have to take into account all these contributions.

Taubes pointed out \cite{Taubes} that the latter contribution to the net potential 
depends on the 
relative orientation of the monopoles in the group space which is parametrised
by an angle $\delta$. Magnetic dipole solution above corresponds to the 
saddle point configuration where the attractive short-range forces,  
mediated both by the $A_\mu^3$ vector boson and the Higgs boson, are balanced by 
the repulsive interaction. The latter forces are mediated by the massive vector bosons  
$A_\mu^\pm$ with opposite orientation 
in the group space. There is a difference from a system of two identically charged BPS 
monopoles where the scalar attraction is cancelled due to contribution of the repulsive gauge interaction.    
Note that the boundary 
conditions on the fields at the spatial asymptotic (\ref{rand}) above, for example for M-A pair $k=1$,   
yield the rotation of the fields on the negative semi-axis $z$ by $\pi$, 
with respect to the fields on the positive semi-axis $z$. 
Evidently, this corresponds to the Taubes conjecture for a magnetic dipole.

\section{Electomagnetic properties of the configurations}
\subsection{Effective electromagnetic interaction} 
  
Thus, the pattern of 
interaction between the monopoles is very different from a naive 
picture of Coulomb long-range electromagnetic interaction of point-like charges. 
It looks a bit surprising, but there is also a possibility to describe this system 
in terms of an effective electromagnetic 
interaction  associated 
with the regular Abelian electromagnetic field strength
tensor 
\begin{equation}  \label{u1tensor}
{\cal F}_{\mu \nu} = \Tr~ \left\{ \hat \Phi F_{\mu\nu} -\frac{i}{2e} \hat \Phi D_\mu \hat \Phi 
D_\nu \hat \Phi \right\} \, .
\end{equation}
Here we make use of the normalization 
$\hat \Phi = \Phi/\eta$. Thus, in a regular gauge all the components of the 
vector field are projected there onto direction of the Higgs field and corresponding 
excitation, a photon, is massive inside of the monopole core. 

This gauge-invariant definition 
of the electromagnetic field strength tensor $F_{\mu\nu}$,  given in \cite{GoddOlive}, 
is close to the original definition of the 't~Hooft tensor \cite{mono}, up to replacement 
$\hat \Phi $ with a normalized Higgs field $\Phi/|\Phi|$. Obviously, both definitions 
coincide on the spacial boundary. 
The difference is that the 't~Hooft tensor 
is singular at the zeros of the Higgs field, while (\ref{u1tensor}) is regular everywhere. 
In both cases the zeros are associated with positions of the monopoles. 

Note that the definition of an electromagnetic field strength tensor is always somewhat arbitrary 
in a non-Abelian gauge theory, for example one can also consider  
${\cal F}_{\mu\nu} = \hat \phi^a F_{\mu\nu}^a$ \cite{GoddOlive,Manton78}. 

Let us briefly recapitulate the electromagnetic properties of the solutions \cite{KKS}. 
The electromagnetic field strength tensor (\ref{u1tensor})  
yields both the electric current $j_{\rm el}^\nu$
\begin{equation}
 \partial_\mu {\cal F}^{\mu\nu} = 4 \pi j_{\rm el}^\nu
\ , \label{jel} \end{equation}
and the magnetic current  $j_{\rm mag}^\nu$
\begin{equation}
 \partial_\mu {^*}{\cal F}^{\mu\nu} = 4 \pi j_{\rm mag}^\nu
\ . \label{jmag} \end{equation}
The magnetic charge of the configuration is defined as 
\begin{equation}
g = \frac{1}{4\pi \eta }\int \frac{1}{2}
{\rm Tr} \, \left( F_{ij} D_k \Phi \right)\varepsilon_{ijk} d^3 r
\ .  \label{magcharge} \end{equation}
In the topologically trivial sector it vanishes, however the charge density distribution 
$g(x) = \frac{1}{2} {\rm Tr} \, \left( F_{ij} D_k \Phi \right)\varepsilon_{ijk}$
is not trivial. Evidently, the electric current $j_{\rm el}^\nu$ 
is vanishing for the spherically symmetric `t Hooft--Polyakov solution. 

As mentioned above, the axially symmetric configurations posess a magnetic dipole moment which, 
in such an effective electomagnetic framework, can be evaluated from the magnetic charge density 
and the electric current density as  \cite{Hindmarsh},
\begin{equation} 
\vec \mu = \left( \mu_{\rm charge} + \mu_{\rm current} \right) \vec e_z
= \int \left( \vec r \, g(x) -
\frac{1}{2} \vec r \times \vec j_{\rm el} \right) d^3 r
\ . \end{equation}
Results of the numerical calculations show that the estimated dipole moment  $\mu_{\rm est}$
quite agrees with the exact value $\mu$ obtained from asymtotic expansion \cite{KKS} (see Table 1)\\ 

\parbox{\textwidth}{
\centerline{
\begin{tabular}{|c|cc|cc|}
 \hline
   \multicolumn{1}{|c|}{}
 & \multicolumn{2}{|c|}{$\mu$}
 &  \multicolumn{2}{|c|}{$\mu_{est}$} \\ 
 \hline
$n/k$ &  1   &   2  &    1 &   2 
 \\
 \hline
1     & ~4.71 & ~9.87~ & ~4.43  & ~9.48~ \\
 \hline
2     & ~4.75 & ~9.63~ & ~5.38 & ~9.76~ \\
 \hline
3     & ~5.20 & ~9.96~ & ~4.92 & ~10.07~ \\
 \hline
4     & ~5.75 & ~10.65~ & ~5.26 & ~10.84~ \\
 \hline
\end{tabular}\vspace{7.mm}}
{\bf Table 1}
The dipole moment $\mu$ and the estimated electromagnetic 
dipole moment $\mu_{\rm est}$ 
are given for the solutions of the 
first and second branches with several values of $n$ at $\lambda = 0$.\vspace{7.mm}\\
}

Thus, the physical picture of the source of the dipole moment is
that it originates both from a distribution of the magnetic charges
and the electric currents \cite{KKS}. Because of axial symmetry of the configurations,
${\vec \mu} = \mu \vec e_z$.

Let us consider the charge and current 
density distributions given by the relations (\ref{magcharge}) and (\ref{jel}), respectively. 
We can evaluate these electromagnetic quantities 
by  straightforward substitution of the numerical 
solutions with given $k$ and $n$ into the definitions above.  

To construct solutions subject to the above boundary conditions,
we map the infinite interval of the variable $r$
onto the unit interval of the compactified radial variable
$x \in [0:1]$,
$$
x = \frac{r}{1+r}
\ , $$
i.e., the partial derivative with respect to the radial coordinate
changes according to
$
\partial_r \to (1- x)^2\partial_{x}
\ . $
The numerical calculations are then performed with the help of the FIDISOL package
based on the Newton-Raphson iterative procedure \cite{FIDI}. 
The equations are discretized on a non-equidistant grid in $x$ and $\theta$
with typical grids sizes of $70 \times 60$.
The estimates of the relative error for the functions
are of the order of $10^{-4}$. The results were presented in \cite{mapKK,KKS}.

Making use of these solutions, we can see that there is a 
single electric current ring for the first branch of the solutions 
with $k=1$
(see Fig. \ref{f-1}). This current is circulating exactly between the maxima of distributions of 
the positive and negative charge density. For $n=1$ the latter are located at the $z$ axis and 
coincide with positions of the nodes of the Higgs  field. For solution with $n=2$ the maxima of the
charge density distribution form two rings parallel to $xy$-plane where the current ring is placed.
Also the energy density of the configuration represents two tori. 
However two (double) zeroes of the Higgs field still are on the $z$ axis. 
The energy density tori are located symmetrically with respect 
to the nodes of the Higgs field. 

\begin{figure}\lbfig{f-1}\begin{center}
  \includegraphics[height=.35\textheight, angle =-90]{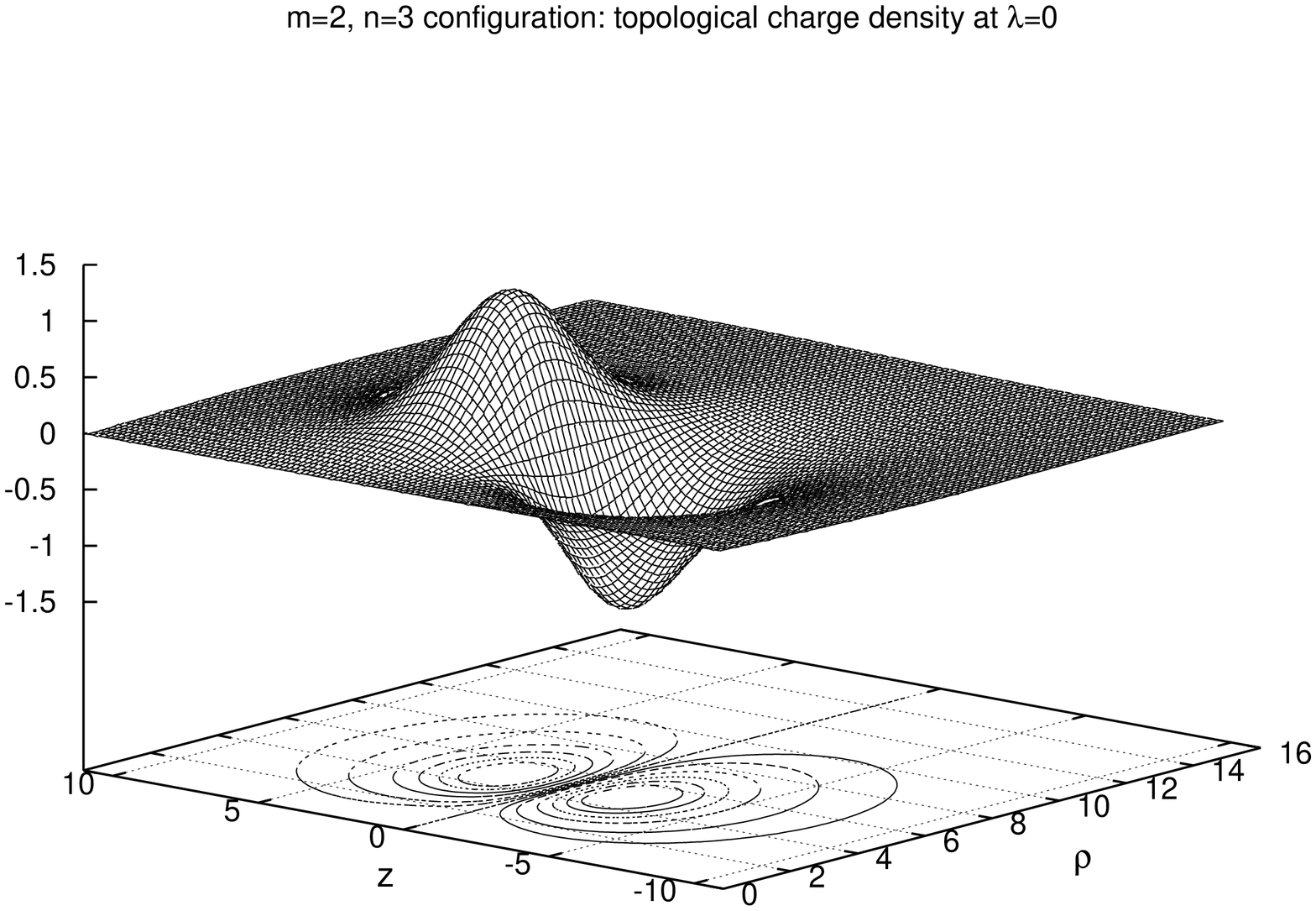}
\includegraphics[height=.35\textheight , angle =-90 ]{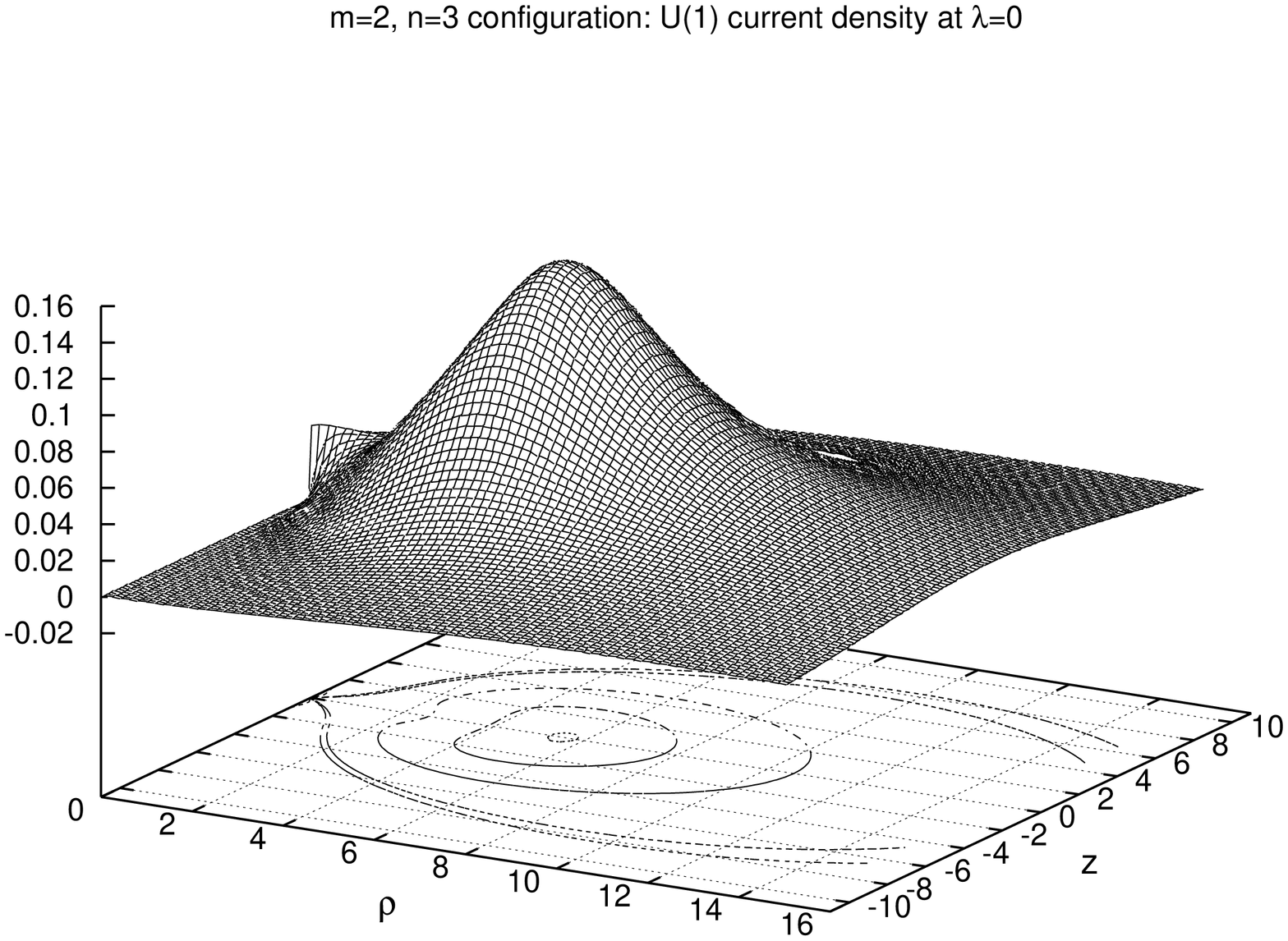}
\end{center}
  \caption{The charge density (left) and electric current density (right) are 
shown for solution with $k=1,n=3$ at $\lambda = 0$ as functions of the 
coordinates $z, \rho$.
}
\end{figure}


\begin{figure}\lbfig{f-2}\begin{center}
  \includegraphics[height=.16\textheight, angle =0]{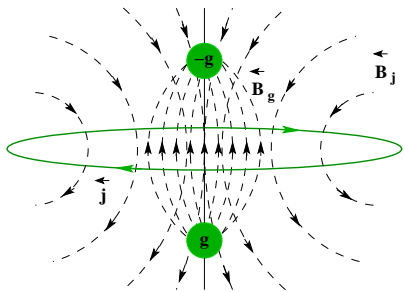}
\end{center}
  \caption{Electomagnetic picture of an 
equilibrium state of two opposite magnetic charges in the magnetic 
field of the electric current ring.
}
\end{figure}


As winding number $n$ increases further, i.e., $n\ge 3$, instead of isolated nodes on the 
symmetry axis, the closed vortex 
solution arises \cite{KKS}. For this configuration  
the Higgs field vanishes on a closed ring centered around the symmetry axis. 
This ring coincides with position of maxima of the 
current density distribution. Thus, 
one may conjecture that the stability of the chain and vortex configurations is  
due to balance of the effective electromagnetic 
interaction between the poles and current rings. Indeed, 
the $U(1)$ magnetic field   
is generated both by the static distribution of the 
density of the magnetic charges, and by the circular 
electric current.  
There is the magnetic field ${\cal B}_k$ of the 
current loop which stabilizes the configuration and counteract the attraction between 
two opposite charges (cf. Fig.~\ref{f-2}). 

For second branch of the solutions with $k=2$ three current rings centered around the symmetry 
axis arise, one counterclockwise 
in the $xy$-plane, and two outer clockwise rings which lie in planes parallel to the $xy$-plane. 
As in the case of the first branch, 
these current rings are circulating between the maxima of distributions of 
the positive and negative charge density. The effect of the magnetic 
field generated by these currents is to provide a balance of electromagnetic interactions in such a 
system of charges and currents. 
  
Let us consider, for example the M-A-M-A chain with $n=1$.  
There are two pairs of maxima of positive and negative charge density distribution on the 
$z$ axis associated with position of the nodes of the Higgs field and the radius of the inner  
current ring is a bit smaller than the radius of two other rings (see Fig.~\ref{f-3}). 

\begin{figure}\lbfig{f-3}\begin{center}
  \includegraphics[height=.35\textheight, angle =-90]{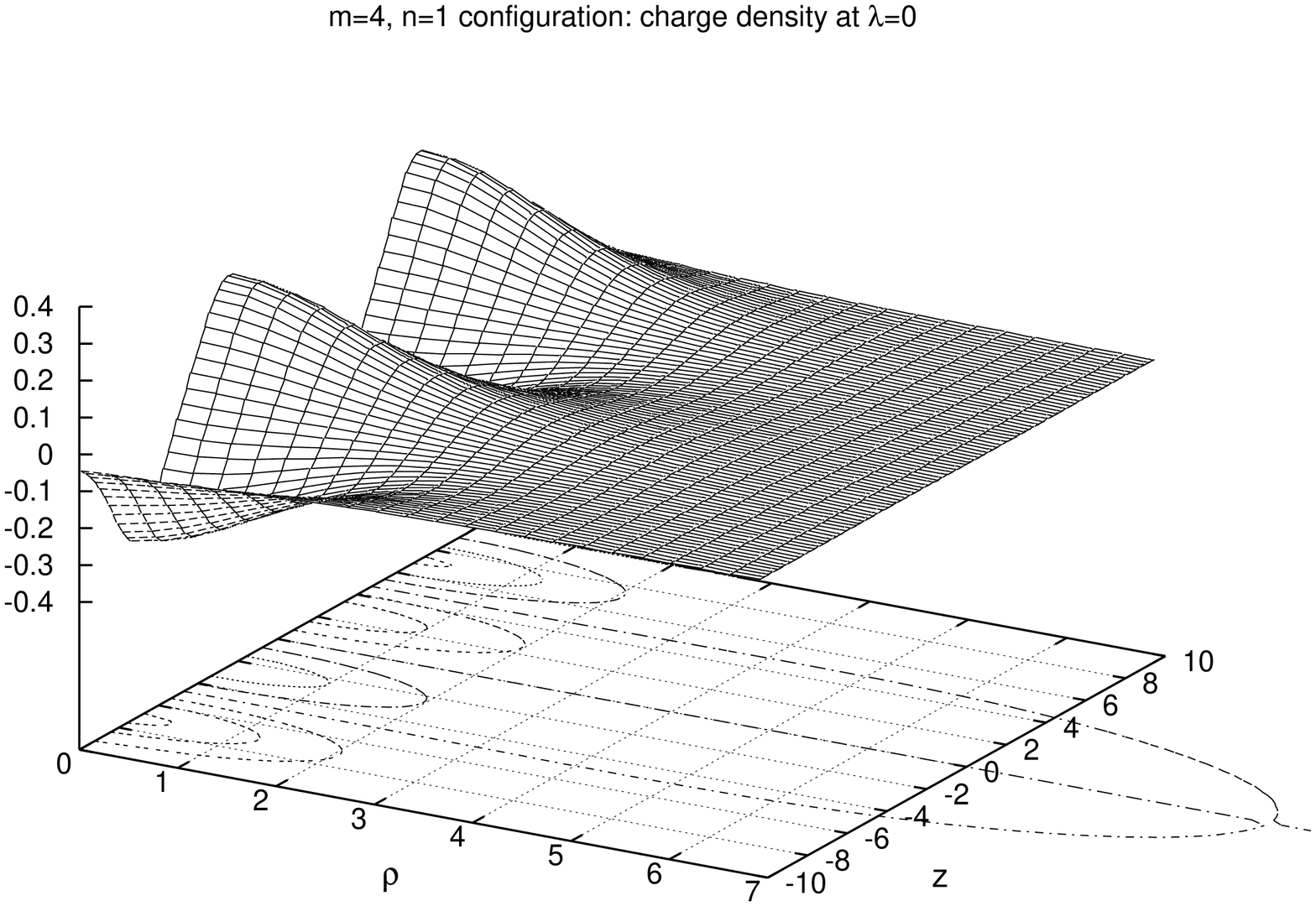}
\includegraphics[height=.35\textheight , angle =-90 ]{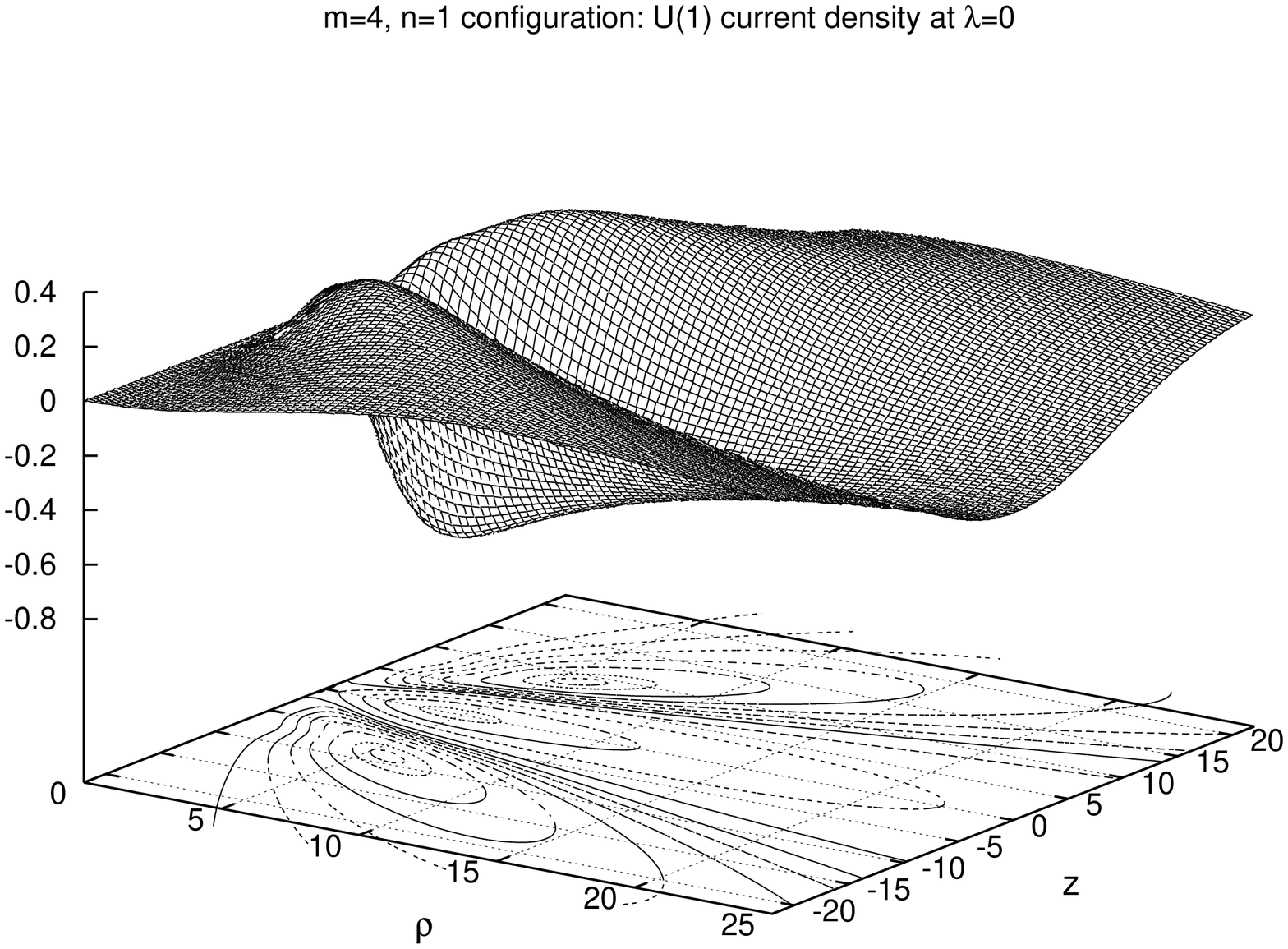}\end{center}
  \caption{The magnetic charge density (left) and the electric current density (right) 
distributions are 
shown for M-A-M-A chain solution with $k=2,n=1$ at $\lambda = 0$ as functions of the 
coordinates $z, \rho$.
}
\end{figure}


As winding number $n$ is increases further, the situation is changing. The maxima of the charge density 
distribution of the $k=2$ configuration with $n=2$ are no longer coincide with double 
nodes of the Higgs field which still are on the $z$ axis. They form four rings with current density tori 
squeezed between them. Further increasement of $n$ leads to separation of the nodes from the symmetry 
axis. As $n$ increases beyond two, the nodes move onto the $\rho$ axis and form two vortex rings located 
symmetrically. The radius of the rings of nodes increases with $n$. Also the radius of three current loops   
increases, however, for the vortex configuration the radius of the inner ring is getting larger than two 
outer rings. The values of the corresponding parameters of all these rings depend on the scalar 
coupling $\lambda$. In general, for larger values of $\lambda$ the radius of the rings is smaller 
and they are closer to the $xy$-plane.

\subsection{Coupling with an external electromagnetic field: Forced axially symmetric 
configurations}

To analyse the structure of the net potential of interaction between the poles, 
let us couple the configurations with an external electromagnetic field.  
This can be done by addition to  the Lagrangian (\ref{lag}) 
a gauge invariant  term 
$$ 
L_{int}= \frac{1}{2}
\varepsilon_{nmk}{\cal F}_{mk} B^{ext}_n \, ,
$$ 
which describes the direct electromagnetic 
interaction between the 
magnetic field of the field configuration and the external homogeneous magnetic 
field $ B_{ext}$  \cite{KS}. To preserve the axial symmetry of the configuration, we suppose 
that this field is directed along $z$ axis, either in positive or in negative direction.  

Such a field can be used as a free perturbation parameter of the model which  
allows to manipulate the 
configuration moving the nodes and transforming its structure. If the forced configuration
remains static,  the energy of the related quasi-elastic deformation is 
balanced by the energy of 
interaction, thus the structure of the net effective potential of the electromagnetic 
interaction may be revealed. 

It was shown 
that for a single `t Hooft-Polyakov monopole inclusion of this term of interaction 
yields an excitation of the 
monopole zero modes and accelerate the monopole precisely 
by analogy with a motion of a point-like charge in 
an external electromagnetic field \cite{KS}. 
Let us consider how inclusion of such a term affects the static axially symmetric 
deformations of the topologically trivial sector. 

The $k=1$ branch begins from the unperturbed magnetic dipole solution $n=1$ \cite{Rueber,mapKK}
with two zeros of the Higgs field on the symmetry axis. 
The charge density distribution has two picks 
of opposite sign, located on the positive and negative $z$ axis, respectively. 
They are associated with 
positive and negative magnetic charges of the dipole \cite{mapKK}. 
In the BPS limit the rescaled mass of configuration is $M_0 = 1.69$ and the distance 
between the nodes is $z_0 = 4.18$ in units of v.e.v. of the scalar field 
\cite{Rueber,mapKK}. As $\lambda$ increases the equilibrium distance is getting smaller 
and for $\lambda = 0.5$ we have  $z_0 = 3.24$ and  $M_0 = 2.48$.

When an external interaction perturbs the configuration, the situation changes. 
Numerical calculations shows that   
if the external magnetic field is directed along negative direction of the $z$ axis 
there is an additional electromagnetic attraction in the system which pushes the poles 
closer to each other. 

We observe similar behavior both for a finite and vanishing scalar coupling $\lambda$. 
This case resembles the situation when gravity is coupled \cite{KK,KKS4}.
As the magnitude of the external magnetic field $B_{ext}$ increases, the nodes of the scalar field, 
which are associated with 
position of the poles, are approaching to each other. The  
potential energy $V_{int}$ of electromagnetic interaction on that branch, 
which is calculated as the difference $V_{int}= E - M_0$, depends almost 
linearly on the external field (see Fig~\ref{f-4}) for positive values of $B_{ext}$. 

\begin{figure}\lbfig{f-4}\begin{center}
  \includegraphics[height=.4\textheight, angle =-90]{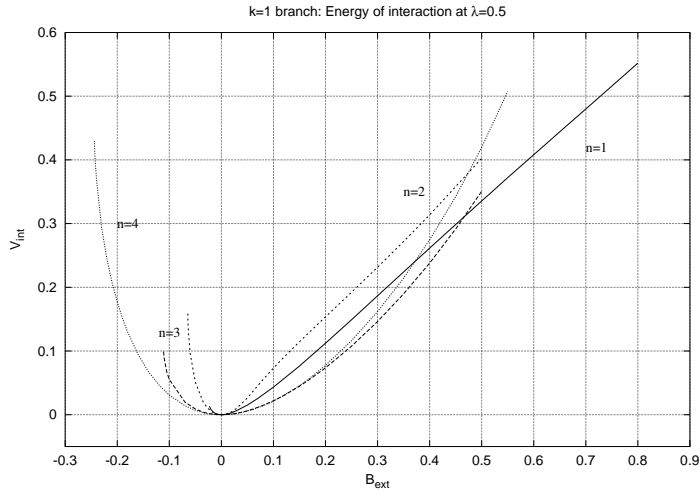}
\end{center}
\caption{Energy of interaction of the static solutions of the $k=1$ branch  
as a function of an external magnetic field at $\lambda = 0.5$.}
\end{figure}


Evidently, this corresponds to the expectable electrodynamical picture of interaction of 
two pointlike opposite charges with an external homogeneous field. As the magnitude of the external 
field increases further, the nodes of the solution continue to approach each other.
However a novelty 
is observed as the external field reaches some critical value, for $\lambda = 0.5$ 
it happens as  $ B_{ext} = 0.515$ and the poles merge at the 
origin. Here the pole and antipole do not annihilate, however. 
The configuration cannot annihilate into the trivial sector 
because of the boundary conditions which we imposed on the spacial infinity. 
Only the modes, that are orthogonal to the Ansatz  (\ref{ansatzA}) may contribute 
to this process.  

As the magnitude of the external magnetic field $ B_{ext}$ becomes stronger,   
a single vortex ring appears instead of a monopole-antimonopole pair. 
The radius 
of the ring is increasing with further increase of the magnitude of the external field. 
However, the energy of interaction still linearly depends on the magnitude of the external  
field.  
\begin{figure}\lbfig{f-5}\begin{center}
  \includegraphics[height=.35\textheight, angle =-90]{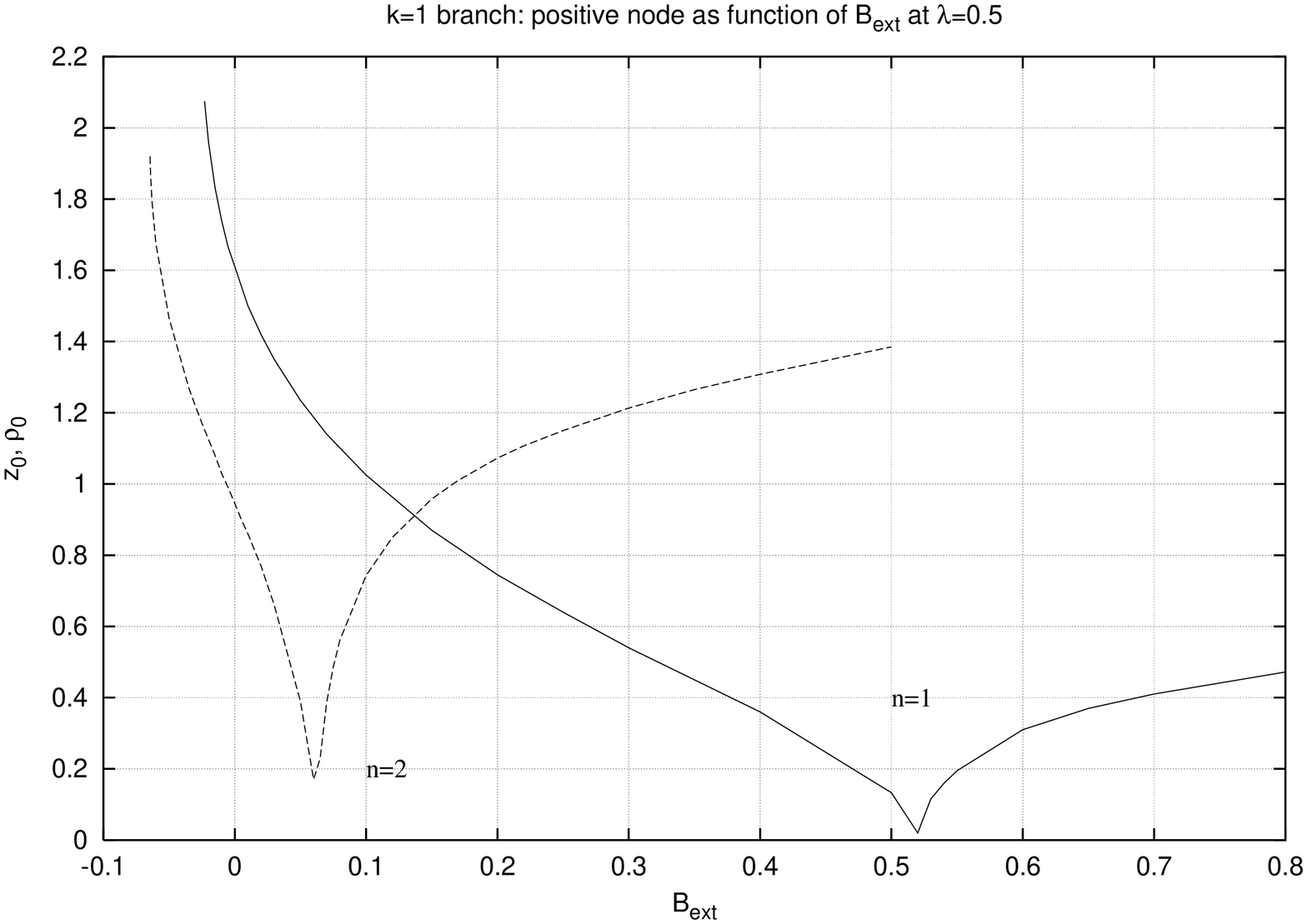}
  \includegraphics[height=.35\textheight, angle =-90]{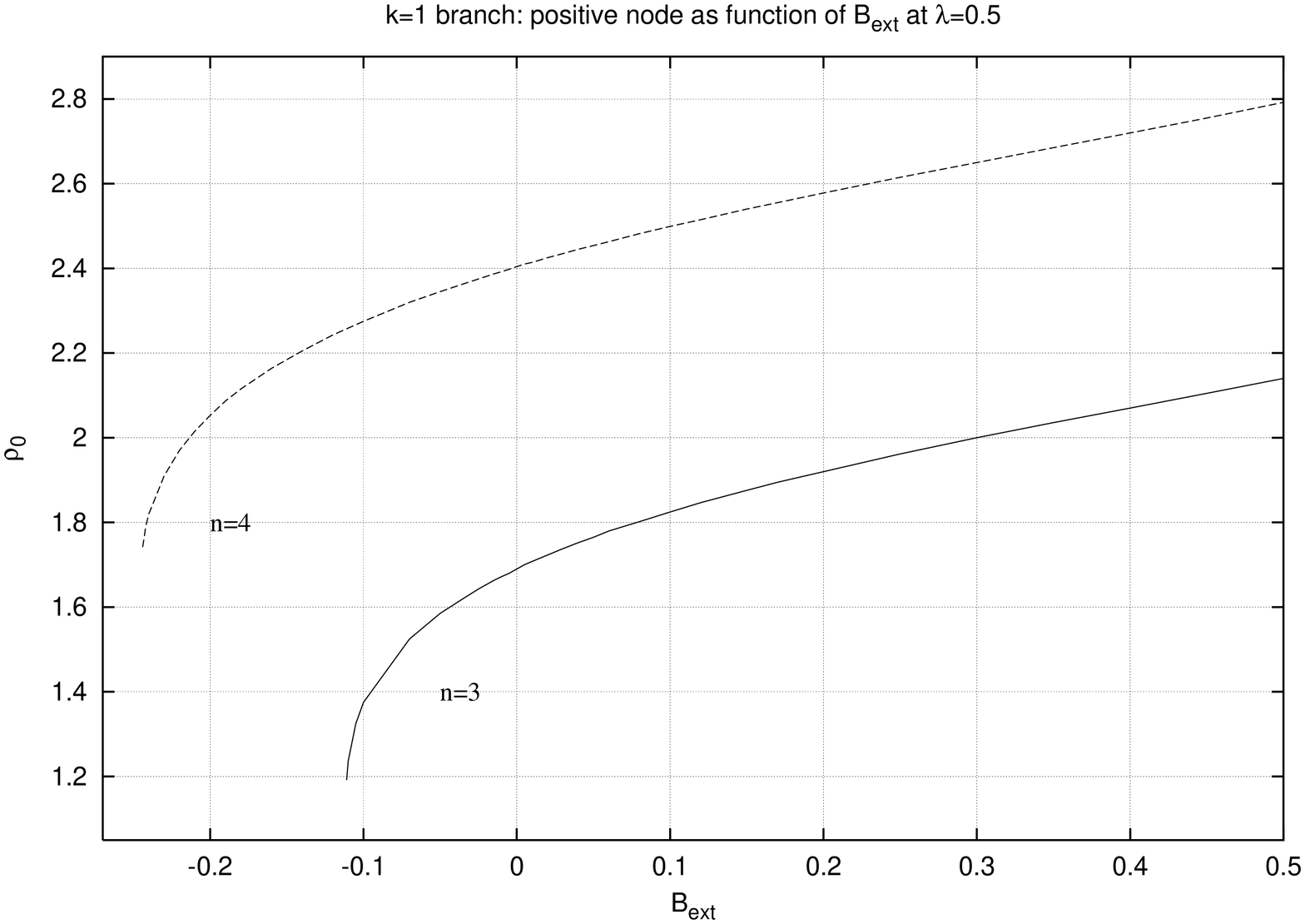}\end{center}
\caption{Position of nodes of the Higgs field of the chain solutions of 
the $k=1$ branch ($n=1,2$ left, and $n=3,4$, right) 
as function of the external magnetic field 
at $\lambda = 0.5$.}
\end{figure}

%

If the external magnetic field is directed along positive direction of the $z$ axis, 
there is an additional repulsion in the system. 
We observed that for a finite value of scalar 
coupling, the potential of electromagnetic interaction develops a local minimum. 
Indeed, in this case with increasing  the magnitude of the 
external magnetic field, the poles moves away from each other 
until some critical distance between the nodes is reached. 
Then the numerical errors are getting relative large and the calculation routine converges
badly. Actually as $\lambda = 0.5$ for $B_{ext} \sim -0.027$ already a 
tiny further increase of the magnitude of perturbation results in large shift of the nodes 
of the scalar field (see Fig.~\ref{f-6a} right). With increasing $\lambda$ the minimum of the 
potential becomes sharper. Considering the solutions in the BPS limit, however, we 
do not find a static monopole-antimonopole pair BPS solution for any finite value of 
$ B_{ext}$.


\begin{figure}\lbfig{f-6a}\begin{center}
  \includegraphics[height=.35\textheight, angle =-90]{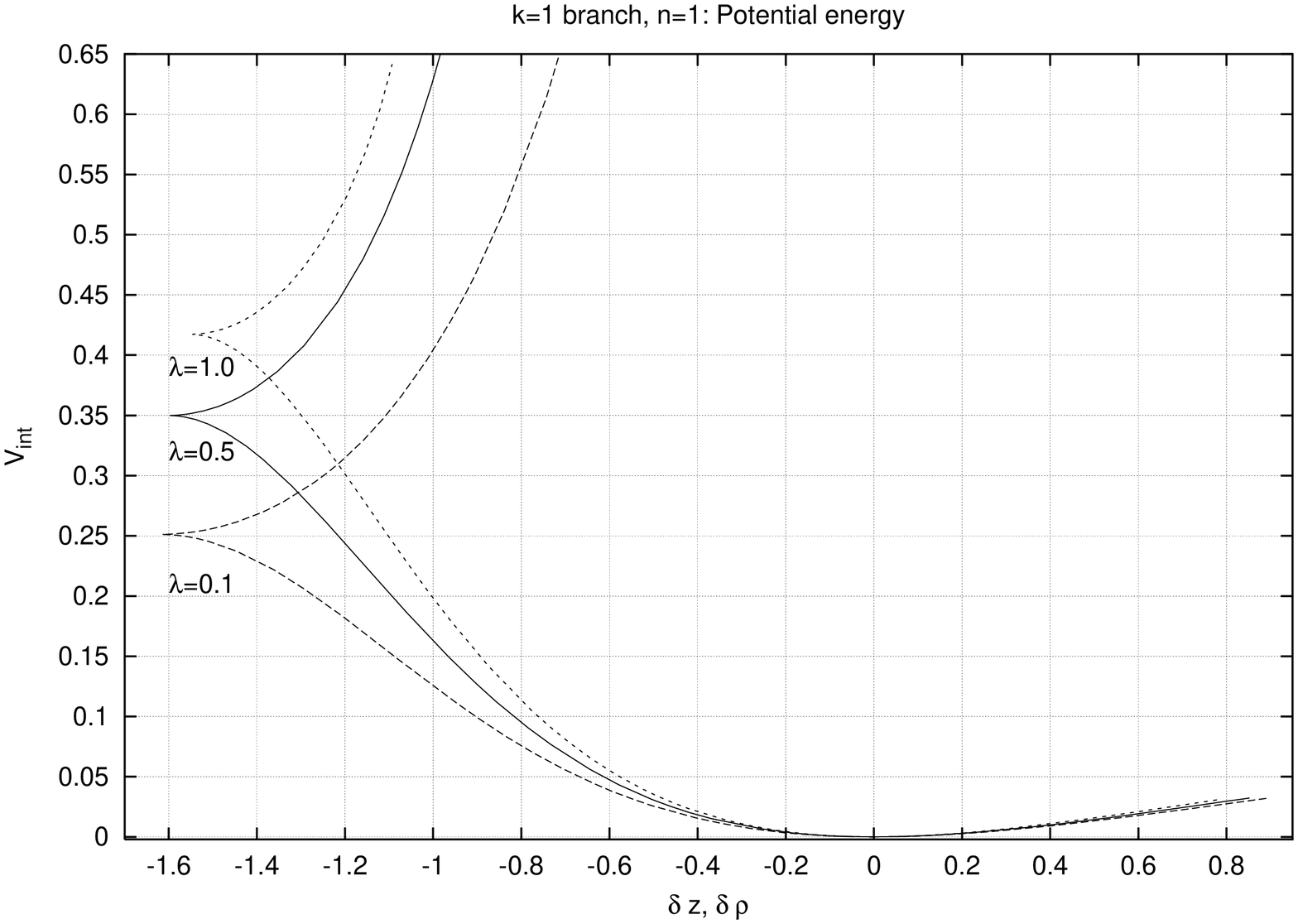}
  \includegraphics[height=.35\textheight, angle =-90]{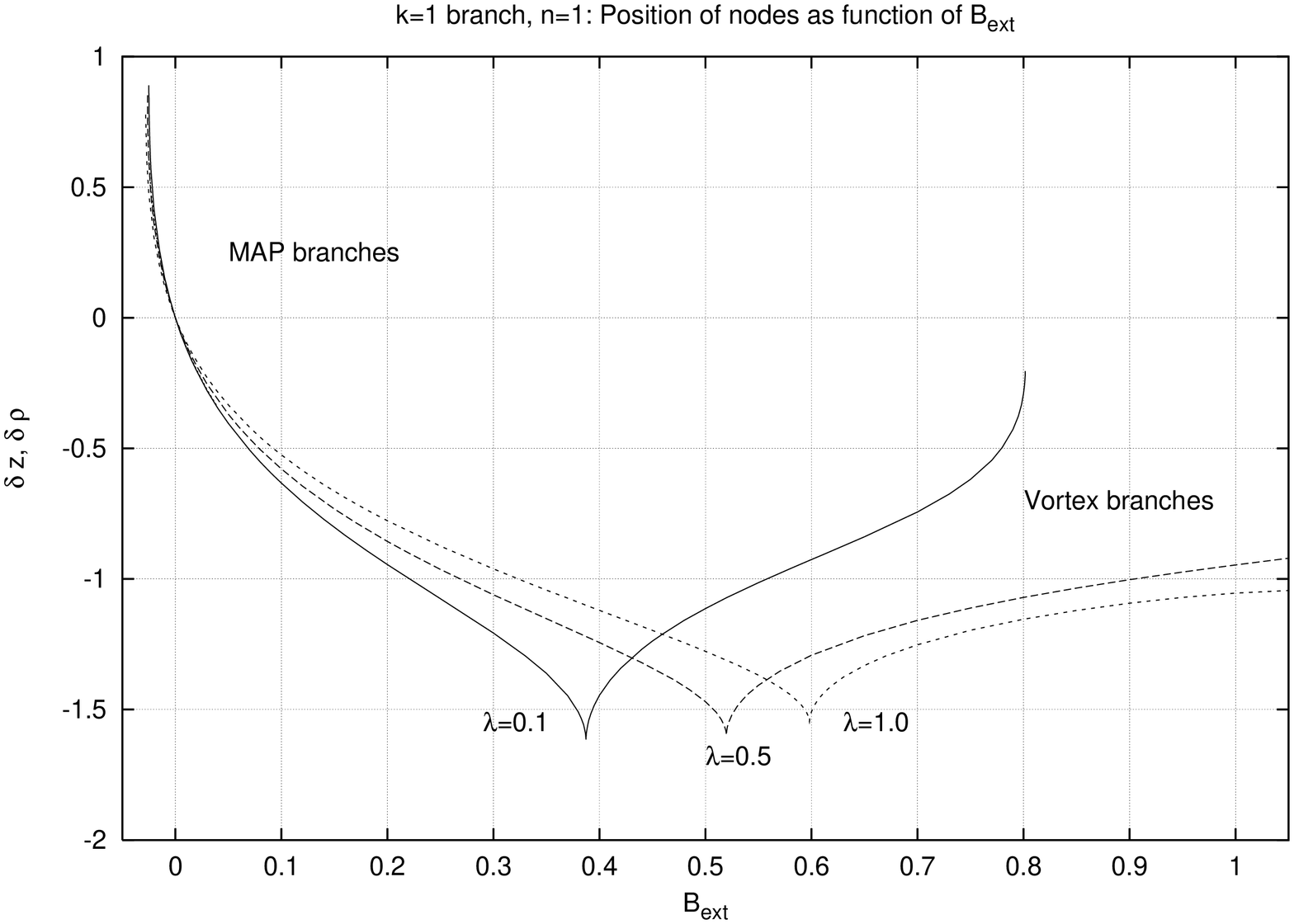}\end{center}
\caption{Potential of monopole-antimonopole interaction 
of the $k=1$ branch  (left) and position of the nodes as a function of $B_{ext}$
(right) for different  values of $\lambda$.}
\end{figure}

We observe a similar scenario considering configuration with $k=1, n=2$. 
Again, for a finite value 
of $B_{ext}$ a branch of solutions emerges from the non-interacting solution. 
However, the energy of interaction 
between the double charged poles in that case is stronger, thus the monopole-vortex transition is
observed at the smaller value of  $B_{ext}$
as in the case of  $k=1, n=1$ configuration. Furthermore, the potential well is also deeper 
(see Fig.~\ref{f-6}).  

Recall that the solutions with  $k=1, n \ge 3$ are no longer characterized by some discrete 
set of isolated poles on the symmetry axis.   
They are the vortex rings already, without an external interaction. The dipole moment of these 
configurations arises solely from the electric current (\ref{jel}) and the current ring appears 
there as a source of the magnetic field \cite{KKS}. 
Testing the vortex solutions with external magnetic field we observe quite 
expectable electrodynamical picture: the radius of the current ring is increasing as  
the magnitude $B_{ext}$ of the magnetic field, which is  
directed along positive direction of the symmetry axis, increases. In contrast, as 
the external field is 
directed in opposite direction, the vortex ring is shrinking.
This branch is 
expended up to some critical value of $B_{ext}$  beyond which external field becomes too strong 
for static configuration to persist. 

Since for a static configuration the energy of interaction with an external field is equal to 
the potential energy of the quasi-elastic deformation, the shape of the potential energy well can be 
easily recovered from the consideration above, see Figs.~\ref{f-6a},\ref{f-6}. 
Evidently, there is a bifurcation which resembles a second order  
phase transition for the solutions of the first branch with $n=1,2$. This bifurcation  
is related with transformation 
of the dipole configuration into the vortex ring. 
With increasing $\lambda$ the potential well is getting more deep.

\begin{figure}\lbfig{f-6}\begin{center}
  \includegraphics[height=.35\textheight, angle =-90]{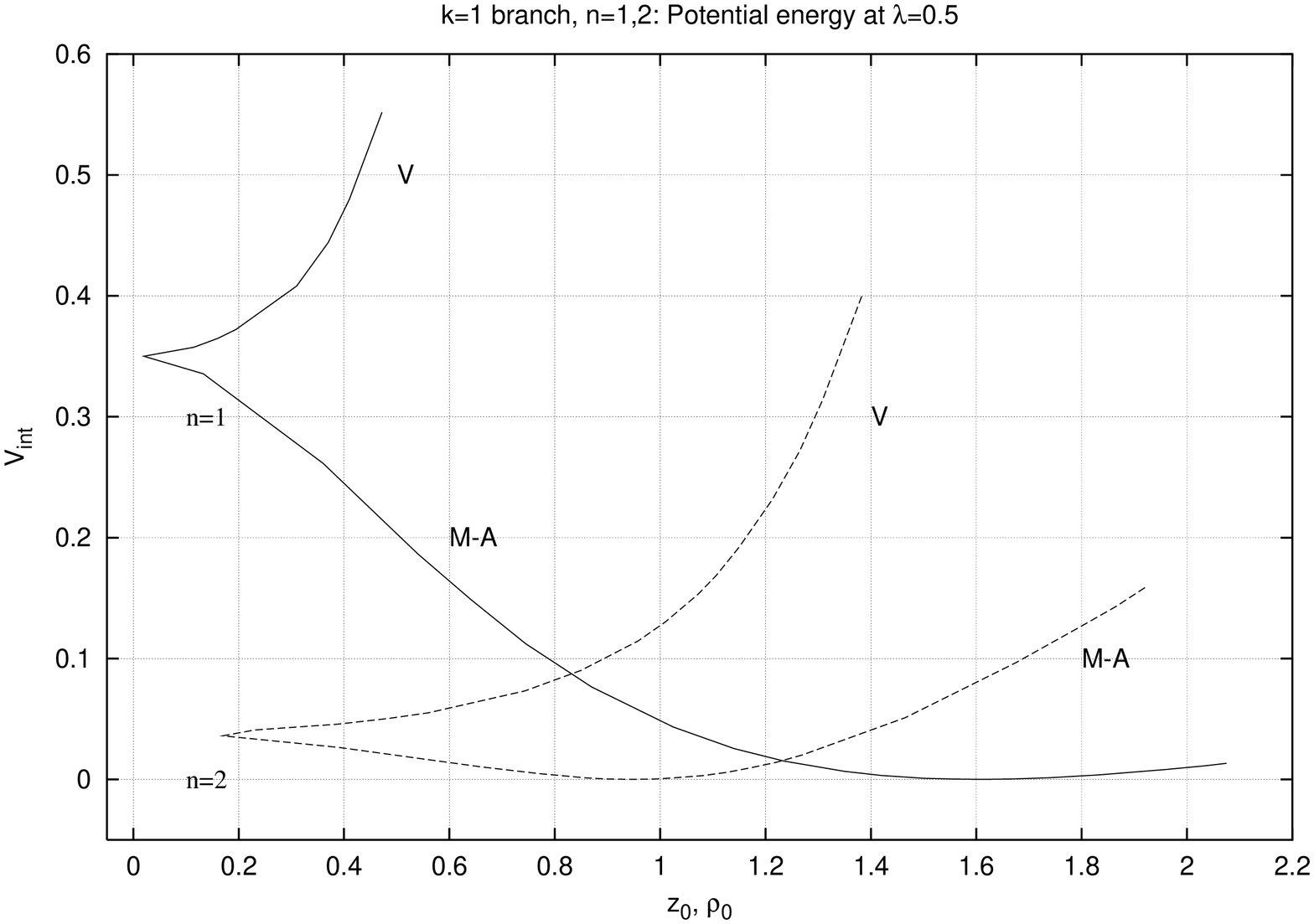}
  \includegraphics[height=.35\textheight, angle =-90]{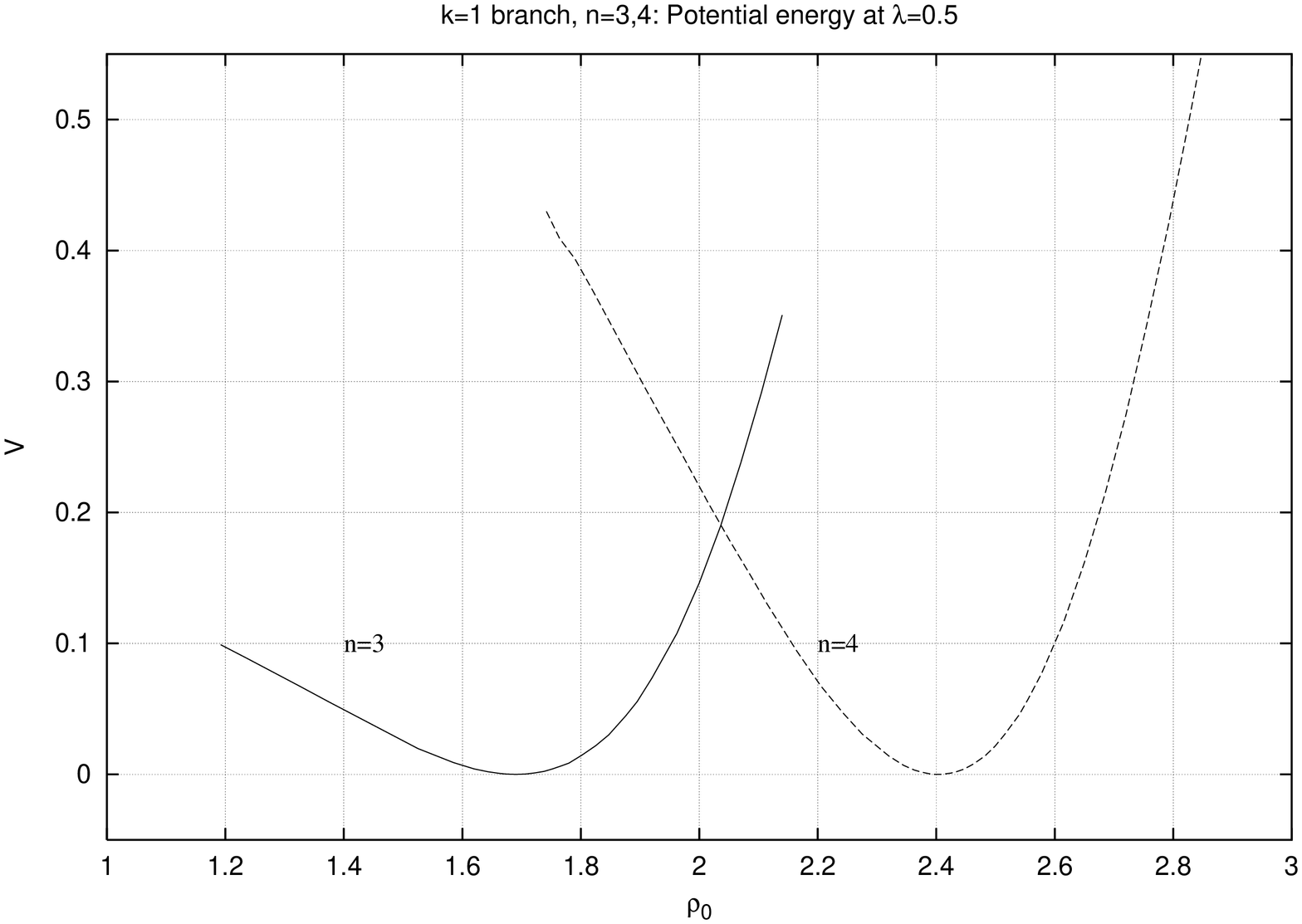}\end{center}
\caption{Potential energy of the static solutions of the $k=1$ branch 
at $\lambda = 0.5$.}
\end{figure}

Similar behavior under external perturbation develop also the solutions in the $k=2$ branch. 
For example,  $n=1$ configuration is a chain solution M-A-M-A, which represent 2 interpolating 
monopole-antimonopole pairs. As magnitude of the external field increases, 
each of two M-A pairs evolves similar to the scenario above; relative potential of interaction
almost exactly reproduces the plot presented in Figs.~\ref{f-6a} and \ref{f-6}. 
Likewise, there is a chain-rings transition for a critical value of the external magnetic
field, each pair forms a vortex ring those radius is increasing as external magnetic 
field becomes stronger. The relative distance between the rings also a bit decreases as
external field is increasing.   

As external magnetic field is directed along positive direction of the $z$-axis, 
the external poles are moving away from each other while the 
location of the inner nodes moves continuously
inward, thus the inner M-A pair is 
squeezing down. There are some indications that for a very large value of the scalar coupling 
this pair merges into a vortex ring. 

Recall that structure of the nodes of the Higgs field
for the solutions with  $n \ge 3$  on that branch is already different:
the modulus of the Higgs field
vanishes in two planes parallel to the $xy$-plane which are
centered around the $z$ axis \cite{KKS}. 
For this double vortex ring solutions, we observe a 
completely analogous pattern 
as for the single vortex on the branch $k=1$: an external electromagnetic interaction 
allows us to manipulate the radius of these rings in a similar way. 

For the $k=3$ branch, we observe the same general pattern again.
The chain solutions ($n \le 2$) consist of 3 monopole-antimonopole pairs and for each pair
the dependence of the location of nodes on the external field reproduces the picture above.
The same pattern holds for vortex solutions:  
for the triple vortex ring solutions  on that branch ($n \ge 3$) each vortex ring evolves as 
a single vortex solution on the branch $k=1$.  
 
\section{Conclusions}
We have shown, that the complicated structure of the short-range Yukawa 
interactions whose balance yields the static axially symmetric non-BPS 
solutions of the $SU(2)$ Yang-Mills-Higgs model, 
may be modelled by an effective potential 
of the electromagnetic interaction between the poles.  
For a given axially symmetrical Ansatz 
there is a local minimum of such a potential 
of interaction. The structure of the electromagnetic interaction in such a system 
may be investigated  
as the configuration becomes coupled with an 
external weak homogeneous magnetic field. The latter is used as a 
perturbation parameter. This perturbation allows us to 
reveal a monopole-vortex transition that is observed at some critical value of the 
magnitude of the external field. 

One may conjecture that 
such a potential of monopole-antimonopole interaction may result, for example,  
in appearance of an intermediate long-living 
breather state in the process of 
annihilation of the configuration into the trivial vacuum. Study of this 
process will be presented elsewhere. 

We believe that the general axially-symmetric Ansatz 
(\ref{ansatzA}) may also describe two 
separated monopoles, which are aligned in the 
isospace. Then it may be possible to provide a similar picture 
of an effective electromagnetic interaction between the monopoles as well. 
By making use of the simple electromagnetic analogy
one may conjecture that there is a current ring associated with each monopole.   
However such a static solution may exist in the BPS limit only. 

\begin{acknowledgments}
I am grateful to  B.~Kleihaus, J.~Kunz 
and P.~Sutcliffe for useful discussions and comments. 
I would like to acknowledge the hospitality at the
Bogoliubov Laboratory of Theoretical Physics, JINR
where this work was completed.
\end{acknowledgments}


\begin{thebibliography}{9}

\bibitem{Manton-book}N.S.~Manton and P.M.~Sutcliffe, {\it Topological Solitons}  
(Cambridge University  Press 2004)
\bibitem{S-book}Ya.M.~Shnir,  {\it Magnetic Monopoles} (Springer,  
Berlin Heidelberg New York 2005)
\bibitem{mono}G.~`t Hooft, 
              Nucl.\ Phys.\ B {\bf 79}, 276 (1974);\\ 
              A.M.~Polyakov, Pis'ma JETP {\bf 20}, 430 (1974).
\bibitem{WeinbergGuth}E.J.~Weinberg and A.H.~Guth,  
                      Phys. Rev.\ D   {\bf D14}, 1660 (1976).
\bibitem{RebbiRossi}C.~Rebbi and P.~Rossi,  
                    Phys. Rev.\ D {\bf 22}, 2010 (1980). 
\bibitem{mmono}R.S.~Ward, 
               Comm. Math. Phys. {\bf 79}, 317 (1981);\\
               P.~Forgacs, Z.~Horvath and L.~Palla, 
              Phys. Lett.\ B {\bf 99}, 232 (1981);\\
               M.K.~Prasad,  
               Comm. Math. Phys. {\bf 80}, 137 (1981);\\
               M.K.~Prasad and P.~Rossi, 
              Phys. Rev.\ D {\bf 24}, 2182 (1981).
\bibitem{monoDS}
see e.g.~P.M.~Sutcliffe, 
        Int. J. Mod. Phys.\ A {\bf 12}, 4663 (1997);\\
         C.~J. Houghton, N.~S. Manton and P.~M. Sutcliffe,  
         Nucl. Phys.\ B  {\bf 510}, 507 (1998).
\bibitem{Rueber} Bernhard R\"uber, Thesis, University of Bonn 1985.
\bibitem{mapKK}
 B.~Kleihaus and J.~Kunz,
Phys.\ Rev.\ D  {\bf 61}, 025003 (2000).
\bibitem{KKS}B.~Kleihaus, J.~Kunz and Ya.~Shnir,
             Phys. Lett.\ B {\bf 570}, 237 (2003);\\ 
              B.~Kleihaus, J.~Kunz and Ya.~Shnir,
Phys.\ Rev.\ D {\bf 68}, 101701 (2003);\\ 
B.~Kleihaus, J.~Kunz and Ya.~Shnir,
Phys.\ Rev.\ D {\bf 70}, 065010 (2004).
\bibitem{jul}
 B.~Julia and A.~Zee,
 Phys.\ Rev.\ D {\bf 11}, 2227 (1975).
\bibitem{wein}
 E.J.~Weinberg,
 Phys.\ Rev.\  D {\bf 20} 936 (1979).
\bibitem{dyonhkk}
 B.~Hartmann, B.~Kleihaus, and J.~Kunz,
 Mod.\ Phys.\ Lett.\ A {\bf 15}, 1003 (2000);\\
B.~Kleihaus, J.~Kunz and Ulrike Neemann, arXiv: gr-qc/0507047.
\bibitem{BPS}E.B. Bogomol'nyi,
              Yad.~Fiz. {\bf 24}, 861 (1976);\\  
               M.K.~Prasad and C.M.~Sommerfeld, 
                Phys.\ Rev.\ Lett.\ {\bf 35}, 760 (1975).
\bibitem{KKT} B.~Kleihaus, J.~Kunz and D.H.~Tchrakian,
              Mod.\ Phys.\ Lett.\ A   {\bf 13} 2523 (1998).

\bibitem{Taubes}C.H.~Taubes,
                  Commun.\ Math.\ Phys.\ {\bf 97}, 473 (1985); ibid 
                  {\bf 86}, 257 (1982); 
                  ibid 
                  {\bf 86}, 299 (1982).
\bibitem{Manton77}  N.S.~Manton, 
                  Nucl.\ Phys.\ B  {\bf 126}, 525 (1977).
\bibitem{KS} V.G.~Kiselev and Ya.~Shnir,  Phys.\ Rev.\ D {\bf 57}, 5174 (1998);\\
Ya.~Shnir,  Mod.\ Phys.\ Lett.\ A {\bf 19}, 287 (2004).
\bibitem{KK}B.~Kleihaus and  J.~Kunz, Phys.\ Rev.\ Lett.\ 
{\bf 85}, 2430 (2000).
\bibitem{KKS4}
 B.~Kleihaus, J.~Kunz, and Ya.~Shnir,
 Phys.\ Rev.\ D {\bf 71}, (2005) 024013.
\bibitem{PT} 
 V.~Paturyan, and D.H.~Tchrakian, 
 J.\ Math.\ Phys.\ {\bf 45}, 302 (2004). 
\bibitem{GoddOlive}P.~Goddard and D.~Olive, Rep.\ Prog.\ Phys.\ {\bf 41}, 1357 
(1978).
\bibitem{Manton78}N.S.~Manton,  Nucl.\ Phys.\ {\bf B135}, 319 (1978). 
\bibitem{Hindmarsh}M.~Hindmarsh and M.~James, Phys. Rev.\ D {\bf 49},
6109 (1994).
\bibitem{FIDI}
 W.~Sch\"onauer, and R.~Wei\ss,
 J.~Comput. Appl. Math. {\bf 27}, 279 (1989);\\
 M.~Schauder, R.~Wei\ss, and W.~Sch\"onauer,
 The CADSOL Program Package, Universit\"at Karlsruhe,
 Interner Bericht Nr. 46/92 (1992).
\end{thebibliography}
\end{document}